\documentclass[aps,prl,twocolumn,epsf,epsfig,amsmath]{revtex4-2}

\usepackage[utf8]{inputenc}
\usepackage{latexsym}
\usepackage{hyperref}
\usepackage{times}
\usepackage{graphicx}
\usepackage{amsmath}
\usepackage{multirow}
\usepackage{amsthm}
\usepackage{dcolumn}
\usepackage{bm}
\usepackage{ textcomp }
\usepackage{color}
\usepackage{amssymb}
\usepackage{xcolor}
\usepackage{easyReview}
\usepackage{mathdots}
\usepackage{float}
\usepackage{lineno}
\usepackage{todonotes}
\usepackage{array}
\usepackage{braket}
\usepackage{orcidlink}

\newcommand{\beq}{\begin{eqnarray}}
\newcommand{\eeq}{\end{eqnarray}}

\begin{document}
\title{Local Current Algebra for the HK Universality Class }
\author{Yuting Bai$^{1}$,\orcidlink{0009-0007-1551-3932}}
\email{yutingb2@illinois.edu}   

\author{Philip W. Phillips$^{1}$}
\email{dimer@illinois.edu}

\affiliation{$^1$Department of Physics and Institute of Condensed Matter Theory, University of Illinois at Urbana-Champaign, Urbana, IL 61801, USA}

\begin{abstract}

We show that a Hamiltonian  in terms of the local real-space currents obeying an $\mathfrak{su}_1(2)$ affine Lie algebra eliminates the non-locality in the  Hatsugai-Kohmoto model for a doped Mott insulator.  We establish this local correspondence through the Bjorken-Johnson-Low prescription for anomalous commutators.   With this result, we show that the charge susceptibility computed from the current Hamiltonian is identical to that with the elemental
Fermionic fields. Consequently,  the HK model is local in
real space, though not in terms of the Fermionic fields, thereby eliminating the key criticism of this model and reinforcing the utility of current algebras for strong interactions.

\end{abstract}
\date{December 2024}


\maketitle

In high-energy physics, the classic paper by Sugawara\cite{sugawara} heralded a shift away from particle fields to currents as the natural coordinates to formulate problems with strong interactions.  This shift is natural because there is no guarantee that the current-carrying degrees of freedom are particulate  when the interactions dominate. In condensed matter, the precursor to the Sugawara work is Tomonaga's\cite{tomonaga}  description of the interacting 1D electron gas in terms of bosonic sound modes composed, which serves as the starting point of bosonization \cite{mattis1965exact,FDMHaldane_1981,coleman,CoadjointOrbitBosonization}.  As currents conserve the number of particles, they are endemic to integrable models such as the 1D Hubbard model which has an $\mathfrak{su}_1(2)$ Kac-Moody algebra\cite{essler} at half-filling. 

In recent years the Hatsugai-Kohmoto model\cite{hk} (see Eq. (\ref{hk})) has emerged\cite{Tenkila2024, twist,hksc,HKnp1,HKnp2,OHK,MaPRB2025,JinchaoRG,ppfailure,compressibility,ppqnh,mai2026twisting,ppz2,bai2025proof,phillips2026mixing,barry,manning2026decomposing,hackner2025solving} 
, as a solvable alternative to the Hubbard model that still captures Mott physics.  In this model, the charge gap arises entirely from the breaking\cite{ppz2,AndersonHaldane} of the $Z_2$ symmetry of a Fermi liquid, thereby producing well-separated lower and upper Hubbard bands.  That the Hubbard and HK models are local in real and momentum space respectively does not pose an insurmountable hurdle as the latter can be deformed\cite{twist} continuously into the former simply by including the momentum mixing that is central to Hubbard physics.  The complexity of this procedure rests solely on the diagonalization of a finite-dimensional matrix arising from hybridizing $2n$ momenta.  As shown previously \cite{twist}, this procedure converges rapidly as $1/n^2$, thereby achieving competitive results compared to real-space cluster methods with modest computational expenditure.

Notwithstanding, it would be advantageous to reformulate HK as a solvable local model. That is, in the spirit of Sugawara's work\cite{sugawara}, we seek a recasting of the HK model in terms of local currents. Using the Bjorken-Johnson-Low prescription \cite{johnson1966current,bjorken1966applications}, we construct a local Hamiltonian with an $\mathfrak{su}_1(2)$ affine Lie algebra that yields the same equations of motion as the HK model. Note in the level-1 algebra, parafermions are absent\cite{cftpdf}.  To buttress our case,  we show that the charge susceptibility\cite{MaPRB2025}, $\chi(q, \omega)$, computed from the currents is identical to that with the elemental Fermionic fields up to corrections $O(q/\pi\rho)$, where $\rho$ is the charge density. These corrections are well controlled in the Dirac limit $\rho,m \to \infty$ while $v_F = \frac{\pi \rho}{m}=\text{Const}$, when the filling sea becomes infinitely deep.  Ultimately, then, the non-Fermi liquid physics of the HK fixed point \cite{JinchaoRG} can be accessed by a local Hamiltonian without losing solvability, a harbinger of the continuous deformation\cite{twist} of HK to Hubbard\cite{twist}.

Our starting point is the 1D HK Hamiltonian\cite{JinchaoRG},
\beq
 H_{\rm HK}= \sum_{ k  \sigma}\epsilon_{ k} n_{ k \sigma} + U\sum_{ k }n_{ k  \uparrow}n_{ k \downarrow}, \label{eq:starting_point}
 \label{hk}
\eeq
which consists of the standard kinetic energy term, $\epsilon_k$ the dispersion, plus a repulsion, $U$, anytime two electrons occupy the same k-state.  In terms of the bare fermions, $c_{ k\sigma}$,  the occupancy, $n_{k\sigma}=c_{k\sigma}^\dagger c_{k\sigma}$.  Once $U$ exceeds the bandwidth, the unperturbed band splits into upper and lower branches, thereby creating a gap at half filling.  The propagating degrees of freedom for the lower and upper bands are $c^{\xi}_{ k\sigma} \equiv (1-n_{ k  \bar{\sigma}})c_{ k\sigma}\equiv P^{\xi}_{ k\bar{\sigma}}c_{ k\sigma}$ and $c^{\eta}_{ k\sigma} \equiv n_{ k\bar{\sigma}}c_{ k\sigma} \equiv P^{\eta}_{ k\bar{\sigma}}c_{ k\sigma}$, respectively.  We have defined energy projectors $P^{\alpha}_{ k\bar{\sigma}}(\alpha = \eta, \xi)$ for the upper and lower Hubbard bands, which satisfy $P^{\alpha}_{ k\sigma}P^{\beta}_{ k\sigma}=\delta_{\alpha \beta}P^{\alpha}_{ k\sigma}$.  For the HK model, all the physics can be formulated entirely in terms of the momentum-based operators $c_{k\sigma}^\alpha$, thereby leading to its solvability.  This is not possible for the Hubbard model as the real-space counterparts are mixed once the hopping is turned on. The commutators
$\left[c^{\xi}_{ k \sigma}, H_{\rm HK}\right] = \epsilon_{ k } c^{\xi}_{ k \sigma}$ and 
$\left[c^{\eta}_{ k \sigma}, H_{\rm HK} \right] = (\epsilon_{ k }+U) c^{\eta}_{ k \sigma}$
yield immediately the two dispersing bands so that in the Heisenberg picture, $c^{\xi}_{ k \sigma}(t)=e^{-i\epsilon_{ k}t}c^{\xi}_{ k \sigma}$ and $c^{\eta}_{ k \sigma}=e^{-i(\epsilon_{ k}+U)t}c^{\eta}_{\bf k \sigma}$.  We assume the dispersion to be quadratic, $\epsilon_k = \frac{k^2}{2m}-\mu$ and consider hole-doping the half-filled system. In such a case, the ground state is singly occupied up to some maximum momentum.  In general, the crossing of the upper and lower bands with the chemical potential will pick out different momenta, $k_\eta$ and $k_\xi$, respectively.  For the case of hole doping, only the lower band will cross the chemical potential. For simplicity, we define $k_{\xi}\equiv k_F$.

Because particle-number and momentum are both conserved, HK physics can be described by particle-hole pairs. We start with operators $\rho_{k \sigma}$ and $\rho^P_{k \sigma}$,  
\beq
\rho_{k \sigma} &=& \sum_q c^{\dagger}_{k+q \sigma}c_{q \sigma} \nonumber \\
\rho^P_{k \sigma} &=& \sum_q k c^{\dagger}_{k+q \sigma}c_{q \sigma},
\eeq
which stand for spatial modulation of particle-number density and momentum density. To simplify notation, we combine particle-number and momentum densities into a two-component vector $\rho_{\mu q \sigma}=(\rho_{q \sigma},\rho^P_{q \sigma})$.

In the Dirac limit where $k_F,m \to \infty$, the only relevant degrees of freedom are those near $k_{\xi}$ and $k_{\eta}$, and $\rho^P_{k \sigma} \simeq k_F \sum_{q}\text{sgn}(q)c^{\dagger}_{k+q \sigma}c_{q \sigma}$.
We will also find it convenient to define the density for left(L) and right(R) movers via $\rho_{\mu q \sigma} = (\rho_{q \sigma}, \rho^P_{q \sigma}) = (\rho_{q R \sigma} + \rho_{q L \sigma}, \rho_{q R \sigma} - \rho_{q L \sigma})$. It is natural to rewrite the dynamics of the HK Hamiltonian in terms of these particle-number conserving composite operators\cite{sugawara1968field}
On the other hand, although we are using the word "current algebra", it is important to distinguish operators $\rho_{q \sigma}$, $\rho_{q \sigma}^P$ defined in our paper, from the N\"other current coupling to the U(1) gauge field. The current of the HK model does not coincide with the momentum density due to nonlocality\cite{MaPRB2025}.

Because the bare electron splits into $c^{\xi}$ and $c^{\eta}$, the density modulation $\rho_{ k a \sigma}$ also has four components: $\rho^{\alpha \beta}_{ k \sigma}=\sum_{\bf q}(c^{\alpha})^{\dagger}_{ k+q a \sigma}c^{\beta}_{ qa\sigma}$, with $\rho_{ ka\sigma}=\sum_{\alpha , \beta}\rho^{\alpha \beta}_{ k a \sigma}$. Furthermore, since the dispersion is approximately linear in the Dirac limit, the current obeys the time evolution
    $\rho^{\alpha \beta}_{ qa\sigma}(t) = e^{iE^{\alpha \beta}_{ qa}t}\rho^{\alpha \beta}_{ qa\sigma}$ where 
\beq
    E^{\alpha \beta}_{ qa} &=& \begin{pmatrix}
        \text{sgn}( a)v_F q & \text{sgn}( a)v_F q-U \\
        \text{sgn}( a)v_F q + U & \text{sgn}( a)v_F q
    \end{pmatrix} \label{eq:EnergyMatrix}.
\eeq
In this matrix Eq. \eqref{eq:EnergyMatrix}, the index $\alpha ,\beta \in \left\{\xi,\eta \right\}$. $\xi(\eta)$ corresponds to 0(1) of row and column index of the matrix Eq. \eqref{eq:EnergyMatrix}. $\text{sgn}( R)=1$ and $\text{sgn}( L)=-1$. That is, these composite operators are also propagating degrees of freedom in the HK model. The ${ q=0}$ component of $\rho^{\xi \xi}_{ q  \sigma}$ counts the number of singly-occupied states $\rho^{\xi \xi}_{ q=0, a\sigma}=\sum_{ k}n_{ k \sigma}(1-n_{ k \bar{\sigma}})$. Therefore, we introduce $n^{\text{O}}_{k a}=(1-n_{ k a \uparrow})(1-n_{ ka\downarrow})$, $n^{\text{S}}_{ ka}=\sum_{\sigma}(1-n_{ k a \sigma})n_{ k a \bar{\sigma}}$ and $n^{\text{D}} = n_{ k a \uparrow}n_{ k a \downarrow}$ , where $n^{\text{O}}$ counts unoccupied states, $n^{\text{S}}$  single occupancy and finally $n^{\text{D}}$  double occupancy.

 Here, we  write an effective theory to capture the IR dynamics of the hole-doping ground state of the band HK model. Although the band HK Hamiltonian has a macroscopic degenerate ground state, its zero-temperature response properties at equilibrium are not determined by an individual ground state but by small fluctuations around the zero-temperature limit of the grand canonical ensemble density matrix, $\rho_{GS}\equiv \lim_{T \to 0}\rho(T)=\lim_{T \to 0}\frac{e^{\beta (\mu N-H )}}{Tr[e^{\beta (\mu N - H)}]}$. Adopting the notation in \cite{hatsugai1992exactly}, we write the ground-state density matrix 
\beq
    \rho_{GS}=\frac{1}{2^N}\sum_{\left\{ \sigma_k \right\}}\ket{\Psi_G;\left\{ \sigma_k \right\}}\bra{\Psi_G;\left\{ \sigma_k \right\}}
\eeq
in terms of ground-state wave functions $\ket{\Psi_G;\left\{ \sigma_k \right\}} = \prod_{\epsilon_k \le \mu} c^{\dagger}_{k \sigma_k} \ket{0} $. Therefore, it is unnecessary to introduce ground-state degeneracy to capture the dynamics of a single density matrix. Since low-energy fluctuations hardly create double occupancy, the operator $\rho^{\eta \eta}$ has no support in this restricted space, and only $\rho^{\xi \xi}$, $\rho^{\eta \xi}$, and $\rho^{\xi \eta}$ enter our IR effective theory.

We next  reformulate the band HK model completely in terms of $\rho^{\alpha \beta}, J^{\alpha \beta}$.  We must first attend to their  commutation relations. In the limit of an infinitely deep filling sea, their commutation relations can not be determined by the canonical commutation relation for $c$ and $c^{\dagger}$\cite{tomonaga,mattis1965exact}.  To circumvent this problem, we adopt the Bjorken-Johnson-Low (BJL) prescription, which relates the matrix elements with the time-ordered correlation function \cite{treiman2014current, bjorken1966applications,johnson1966current},
\beq
     T(\omega)&=& \int dt e^{i \omega t} \bra{a}TA(t)B(0)\ket{b} \nonumber \\
    \lim_{\omega \to \infty} \omega T(\omega)&=&i \bra{a}[A(0),B(0)]_{-} \ket{b}, \label{eq:BJL},
\eeq
to evaluate the anomalous commutator of this interacting theory. A proof of the BJL prescription is presented in Appendix ~A. This method works equally well for free and interacting systems and can be applied in arbitrary dimensions. 

Since HK's response properties are determined by $\rho_{GS}$, we evaluate  
\beq
    \braket{\rho^{\alpha \beta}_{ q a \sigma}\rho^{\gamma \delta}_{q'b\sigma'}} \equiv \text{Tr}[\rho_{\text{GS}}\rho^{\alpha \beta}_{ q a \sigma}\rho^{\gamma \delta}_{ q' b \sigma'}], 
    \eeq
    to compute the time-ordered current-current correlator from the BJL prescription.  
In the Dirac limit, $k_F, m \to \infty$ with $v_F = \frac{k_F}{m}=\text{const}$, replacing the commutators with their vacuum expectation value is sufficient to describe the IR dynamics. This will be the main approximation in our calculations.


The BJL prescription, Eq.\eqref{eq:BJL}, relates the equal-time commutator of $\rho^{\alpha \beta}_{\mu { q \sigma}}$ with the time-ordered correlation function $T^{\alpha  \beta ;\gamma \eta}_{\mu \nu, \sigma \sigma'}(q, \omega)$,
\beq
    T^{\alpha \beta; \gamma \eta}_{\mu \nu,\sigma \sigma'}(q,\omega )\equiv \int_{-\infty}^{+\infty}dte^{i \omega t}\braket{T\rho^{\alpha \beta}_{ \mu p \sigma}(t)\rho^{\gamma \eta}_{ \nu q \sigma'}(0)} \delta_{ p+q ,0}. \label{eq:T_factor}\nonumber\\
\eeq
Since $\rho_{GS}$ preserves translation and $SU(2)$ spin symmetry, $T^{\alpha \beta ; \gamma \eta}_{\mu \nu;\sigma \sigma'}$ is nonzero only when $\sigma = \sigma'$ and $p+q=0$. Moreover, it is nonvanishing only when $\alpha = \eta$ and $\beta = \gamma$. This can be proven by rewriting $\rho^{\alpha \beta}_{\mu q \sigma}$ in terms of fermion operators. For the $\mu = 0$ component (the density) of the current,
\beq
    \braket{\rho^{\alpha \beta}_{0, q \sigma}\rho^{\gamma \eta}_{0, -q \sigma}}&=&\sum_{ k }\braket{c^{\alpha \dagger}_{ k+q \sigma}c^{\beta}_{ k \sigma}c^{\gamma \dagger}_{ k \sigma}c^{\eta}_{ k+q \sigma}} \\
    &=&\sum_{ k} \braket{P^{\alpha}_{ k+q \bar{\sigma}}P^{\beta}_{ k \bar{\sigma}}P^{\gamma}_{ k \bar{\sigma}}P^{\eta}_{ k+q \bar{\sigma}}n_{ k+q \sigma}(1-n_{ k \sigma})}. \nonumber
    \eeq
Because the projector $P^{\alpha}_{ k\sigma}P^{\beta}_{ k\sigma} = \delta_{\alpha \beta}P^{\alpha}_{ k \sigma}$, we may conclude that $ \braket{\rho^{\alpha \beta}_{ q \sigma}\rho^{\gamma \eta}_{ -q \sigma}} \propto  \delta_{\alpha \eta}\delta_{\beta \gamma}$ when $q \ne 0$. Thus, the only independent nonvanishing components of $T^{\alpha \beta ; \gamma \eta}_{\mu \nu;\sigma \sigma'}$ are $T^{\xi \xi; \xi \xi}_{\mu \nu\sigma}(q,\omega) $ and $T^{\xi \eta; \eta \xi}_{\mu \nu\sigma}(q, \omega)$. 

At first glance, it seems that when $U$ is large, it suffices to consider the projection of the current operator to the LHB, $\rho^{\xi \xi}$. As shown in the calculations in Appendix~B, in the limit $\omega \to \infty$ following with $q \to 0$, we find that the residues of the simple pole $\frac{1}{\omega}$ are
\beq
    T^{\xi \xi; \xi \xi}_{00,\sigma}&=&T^{\xi \xi; \xi \xi}_{11,\sigma} = 0,\\
    T^{\xi \xi; \xi \xi}_{01,\sigma}&=&\frac{qL}{2\pi i \omega} .
\eeq
We may then derive the equal-time commutation relation by applying the BJL prescription Eq.\eqref{eq:BJL},
\beq 
\left[\rho^{\xi \xi}_{ qa \sigma},\rho^{\xi \xi}_{ q'b \sigma'}\right]&=&-\frac{1}{2}\delta_{ ab}\delta_{ q+q',0}\delta_{\sigma \sigma'}\frac{qL}{2\pi} .
\eeq
The principal point here is that we are now in possession of the key building block
of any theory of currents.  Further, since we used the BJL prescription, we have constructed the equal-time commutator from microscopics rather than from phenomenology.

It turns out that including $\rho^{\xi \xi}$ alone is not enough to preserve the conservation law in the IR theory. One must introduce $\rho^{\xi \eta}$ and $\rho^{\eta \xi}$. This is because the projected charge, $N_{L}=N-\sum_{ k }n_{ k \uparrow}n_{ k \downarrow} \equiv N-N_D$, is not conserved. Consider the HK Hamiltonian perturbed by an onsite potential, $H =H_{HK}+\sum_{x\sigma}\rho_{\sigma }(x)V(x)= H_{HK}+\sum_{ k \sigma}V_{ -k}\rho_{ k \sigma}$, with $V_k$ being the Fourier component of $V(x)$. That the projected charge is not conserved can be explicitly derived from the band HK Hamiltonian, which is
\beq
    [N_L, H] &=& -[N_D,H] =-\sum_{\bf k  \sigma}V_{ -k  \sigma }(\rho^{\eta \xi}_{ k  \sigma}-\rho^{\xi \eta}_{ k  \sigma}).
    \label{eq:chargeconservation}
\eeq

The same conservation law must hold for the IR theory. This not only enforces the inclusion of $\rho^{\xi \eta}$ and $\rho^{\eta \xi}$, but also poses constraints on the commutation relation between $\rho^{\xi \xi}$, $\rho^{\xi \eta}$, and $\rho^{\eta \xi}$. Recall that, in terms of current operators, $\rho_{ q \sigma} = \sum_a(\rho^{\xi \xi}_{ qa \sigma}+\rho^{\xi \eta}_{ qa \sigma}+\rho^{\eta \xi}_{ qa \sigma})$. The simplest way to satisfy the conservation law in Eq. \eqref{eq:chargeconservation} for an arbitrary external potential is to maintain that
\beq
\left[\rho^{\xi \xi}_{ qa},\rho^{\xi \eta}_{ q'a'}\right]&=&\delta_{ aa'}\rho^{\xi \eta}_{ q+q' a} \label{eq: LL-LU}\\
    \left[\rho^{\xi \xi}_{ qa},\rho^{\eta \xi}_{ q'a'}\right]&=&-\delta_{ aa'}\rho^{\eta \xi}_{ q+q'a} \label{eq: LL-UL},
\eeq
where spin indices are omitted since currents with opposite spin commute. The fact that one has to introduce extra degrees of freedom $\rho^{\xi \eta}$ and $\rho^{\eta \xi}$ is related to spectral weight transfer when doping a Mott insulator \cite{Tenkila2024,sawatzky}, which implies that projecting to the LHB leads to charge-nonconservation as pointed out previously\cite{Sawatzkyprl}.


We next evaluate the commutator between $\rho^{\xi \eta}$ and $\rho^{\eta \xi}$. Following Appendix~B, one has,
\beq
    T^{\xi \eta; \eta \xi}_{00,\sigma}&=&T^{\xi \eta; \eta \xi}_{11,\sigma}=\frac{1}{-i\omega}\left(\frac{N}{4}-\frac{|q|L}{8 \pi} \right) \label{eq: singular_contribution}\\
    T^{\xi \eta; \eta \xi}_{01,\sigma}&=&T^{\xi \eta; \eta \xi}_{10,\sigma}=\frac{qL}{-8\pi i \omega}.
\eeq
Then, through directly applying the BJL prescription, Eq \eqref{eq:BJL}, we arrive at
\begin{widetext}
    \beq
\left[\rho^{\xi \eta}_{ qb \sigma},\rho^{\eta \xi}_{ q'a \sigma '}\right]&=&\frac{1}{2}\delta_{ ab}\delta_{\sigma \sigma '}\left(\rho^{\xi \xi}_{ qa}+\delta_{ q+q',0}\text{sgn}({ a})\frac{qL}{4 \pi} \theta(\text{sgn}({ a})q)\right).
\eeq
\end{widetext}

We notice that the HK interaction modifies the current algebra in two ways. First, it splits the electron $c$ into $c^{\eta}$ and $c^{\xi}$, thereby leading to an $\mathfrak{su}(2)$ structure in the charge sector of the current algebra by defining the comoponents of the current to be ${\bf \varrho }^{z}_{ qa \sigma}=\rho^{\xi \xi}_{ qa\sigma}$, ${\bf \varrho }^{-}_{ qa \sigma}=2\rho^{\eta \xi}_{ qa \sigma}$ and ${\bf \varrho }^{+}_{ qa \sigma}=2\rho^{\xi \eta}_{ qa \sigma}$.   Next, the appearance of $|q|$ in Eq. \eqref{eq: singular_contribution} leads to a step function $\theta(q)$ in the current algebra, which implies a nonlocal equal-time commutation relation. This is the remnant of the nonlocal nature of the HK interaction $Un_{ k \uparrow}n_{ k \downarrow}$. 

We now show that the dynamics of the band HK model can be captured by a local $\mathfrak{su}(2)$ current algebra  up to corrections of $O(\frac{q}{k_F})$. We construct this local IR theory by keeping the $\mathfrak{su}(2)$ structure of the current algebra of the HK model,
\beq
    \left[\rho^{\xi \xi}_{ qa},\rho^{\xi \xi}_{ q'a'}\right]&=&-\frac{\delta_{{ aa'}}}{2}\text{sgn}({ a})\delta_{ q+q',0}\frac{qL}{2\pi} \\
    \left[\rho^{\xi \xi}_{ qa},\rho^{\xi \eta}_{ q'a'}\right]&=&\delta_{ aa'}\rho^{\xi \eta}_{ q+q' a} \\
    \left[\rho^{\xi \xi}_{ qa},\rho^{\eta \xi}_{ q'a'}\right]&=&-\delta_{ aa'}\rho^{\eta \xi}_{ q+q'a} \\
    \left[\rho^{\eta \xi}_{ qa},\rho^{\xi \eta}_{ q'a'}\right] &=&-\frac{\delta_{ aa'}}{2} \left ( \rho^{\xi \xi}_{ q+q'a}+\frac{1}{2}\text{sgn}({ a})\delta_{ q+q',0}\frac{qL}{2\pi} \right ), \nonumber \\
\eeq
while replacing the singular part $\theta(\text{sgn}({\bf a})q)$ with unity.
Here we have suppressed the spin indices. After defining ${\bf \varrho }^{\pm}_{qa\sigma}={\bf \varrho }^x_{qa\sigma} \pm i {\bf \varrho }^y_{qa\sigma}$, we find that these currents form an affine $\mathfrak{su}_1(2)$ Lie algebra, 
\beq
    \left[{\bf \varrho }^{i}_{ qa \sigma},{\bf \varrho }^j_{ q'a'\sigma '}\right]=\delta_{ aa'}\delta_{\sigma \sigma '}\left(i\epsilon_{ijk}{\bf \varrho }^{k}_{ q+q' a}-\frac{\text{sgn}({ a})}{2}\delta_{ q+q',0}\frac{qL}{2\pi} \right) \label{eq:commutator} \nonumber \\
\eeq
for $i=x,y,z$. This is identical in form to the current algebra of the Hubbard model at half filling \cite{essler}.

It is now easy to build the Hamiltonian $H(\rho^{\alpha \beta})$ (in the Sugawara form \cite{sugawara1968field}) that yields the correct energy $E^{\alpha \beta}_{\bf qa\sigma}$ in the Dirac limit,
\beq
    H = v_F \frac{2 \pi}{L}\sum_{i=x,y,z}\sum_{ qa \sigma}{\bf \varrho }^{i}_{ qa \sigma}{\bf \varrho}^{i}_{ -qa \sigma} - U\sum_{ a  \sigma}{\bf \varrho }^{z}_{ q=0 a 
     \sigma}. \label{eq:Hamiltonian}
\eeq
Here, the first term, which is proportional to ${\bf \varrho }^2$, generates the linear dispersion $v_Fq$, and the second term is analogous to a Zeeman term that splits the energy between $\rho^{\xi \xi}$ and $\rho^{\xi \eta}, \rho^{\eta \xi}$ by an amount $\pm U$. 

We next define the Hilbert space of the Hamiltonian. We start with the vacuum state $\ket{0}_B$, which satisfies
\beq 
    \rho^{\xi \xi}_{ q>0 R \sigma} \ket{0}_B &=& \rho^{\xi \xi}_{ q<0 L \sigma} \ket{0}_B = 
    \rho^{\xi \eta}_{ q a \sigma} \ket{0}_B = 0, \nonumber \\
    \rho^{\eta \xi}_{ q>0, a \sigma} \ket{0}_B &=& \rho^{\eta \xi}_{ q<0, a \sigma} \ket{0}_B = 0,\nonumber \\
    \rho^{\xi \xi}_{ q=0,R \sigma} \ket{0}_B &=&\rho^{\xi \xi}_{ q=0,L \sigma} \ket{0}_B = \frac{N }{4} \ket{0}_B \label{eq:vaccumstatedef}
\eeq
and define the Hilbert space $\mathcal{H}_B$ as the Fock space spanned by $\prod_{i}\rho^{\alpha_i \beta_i}_{q_i a_i \sigma_i} \ket{0}_B$. The action of $\rho^{\xi \xi}_{ q=0,R/L \sigma}$ on $\ket{0}_B$ is counting the average left/right-moving particles with spin $\sigma$ on all HK degenerate ground states. Here, it is worth noticing that each state in the bosonic Hilbert space corresponds to a density matrix in the original fermionic picture. State $\ket{\psi_B} \in \mathcal{H}_B$ corresponds to the density matrix $\hat{\rho}_F$ for the original fermionic system such that for observable $O$, $\bra{\psi_B}O\ket{\psi_B}=\text{Tr}[\hat{\rho}_F O]$, at least up to corrections that scale as $\frac{q}{k_F}$ and $\frac{\omega }{U}$. This correspondence allows our current algebra Hamiltonian to capture the dynamics of the band HK model near equilibrium without introducing ground state degeneracy. 

It is worth noticing that the Hamiltonian Eq. \eqref{eq:Hamiltonian} is local in the currents, which we confirm after performing the Fourier transform $\rho^{\alpha \beta}_{ q a \sigma}=\int dx e^{-i q  x}\rho^{\alpha \beta}_{ a\sigma}(x)$, 
\beq
    H =2 \pi v_F\sum_{ a \sigma} \int dx \vec{\bf \varrho }_{ a \sigma}^2(x)-\sum_{ a \sigma}\int dx U {\bf \varrho }^{z}_{ a \sigma}(x).
\eeq
Hence, we have shown that the dynamics of current in the HK model can be reproduced by a local Hamiltonian. This is consistent with the nonlocal nature of HK interaction because the operators $\varrho^i_{ a\sigma}(x)$ are nonlocal combinations of the elemental Fermionic fields, $c$, $c^{\dagger}$. 

A key element of this Hamiltonian is that the algebra of the charge sector is now $\mathfrak{su}(2)$ instead of $\mathfrak{u}(1)$. The $\mathfrak{su}(2)$ structure arises from the fact that one would like to work with projected currents $\rho^{\xi \xi}$ and satisfy the charge conservation law, as shown in Eq. \eqref{eq:chargeconservation}. This $\mathfrak{su}(2)$ structure is associated with a geometric phase, which could be read off by allowing the second term to couple to all components of ${\bf \varrho}^i_{a \sigma}(x)$,
\beq
    H[{\bf U}(t)] = \sum_{\bf a \sigma} \int dx [2 \pi v_F \vec{\bf \varrho}_{\bf a \sigma}^2(x) - {\bf U}(x,t) \cdot {\bf \varrho}_{\bf a \sigma}(x)]\label{eq:CurrentHamiltonian}.
\eeq
The equation of motion in the Heisenberg picture is thus
\begin{align}
    -i\frac{d}{dt}{\bf \varrho}^{i}_{ qa \sigma}=\epsilon_{ qa}{\bf \varrho}^{i}_{ qa\sigma} + i\epsilon^{ijk}U^j{\bf \varrho}^{k}_{ qa\sigma},
\end{align}
where the second term describes the precession under a magnetic field.  A periodic modification of ${\bf U}(t)$ would lead to a Berry phase. If one were to integrate out the current to arrive at an effective action of ${\bf U}$, this geometric phase would induce a Wess-Zumino-Witten (WZW) term \cite{MStoneBerryPhase}, which forbids the effective action from being a single-valued function of ${\bf U}$. As pointed out previously, the efficient cause is the breaking of the $Z_2$\cite{ppz2} symmetry of the HK interaction.

In terms of the original fermions, $U^x_{ -q}\varrho^x_{ qa \sigma} + U^y_{ -q}\varrho^y_{ qa\sigma}$ is given by
\begin{align}
    (U_x +iU_y)\rho^{\xi \eta}_{ qa\sigma}\Leftrightarrow\sum_{k}(U_x+iU_y)(c^{\xi})^{\dagger}_{ k+q a\sigma}c^{\eta}_{ ka\sigma},
\end{align}
which pushes electrons from the LHB to the UHB with a momentum shift $q$. This makes it explicit that it is the mixing between the bands that produces the non-locality in terms of the original Fermionic operators.  It is worth noticing that directly setting $q$ to zero makes this term vanish.  This further shows that it is the spectral weight transfer between the UHB and LHB that ultimately leads to the anomalous WZW term.  Thus, to obtain the Berry phase, one must consider $H[{\bf U}(x,t)]=H[{\bf U}(t)+{\bf \epsilon u}(x,t)]$ with infinitesimal nonuniformity, and take the limit $\epsilon \to 0$.

We next show that the correlation function of our current algebra Hamiltonian matches with the band HK model up to corrections of $\frac{q}{k_F}$.
\beq 
    \chi(q,\omega)=\left(\rho - \frac{\lvert q \rvert }{2 \pi} \right) \frac{U}{\omega^2 - U^2}+\frac{v_F q^2}{ \pi \omega^2} + o_{q \to 0}(q^2) \label{eq:chi_HK},
\eeq
with $\rho = \frac{N}{L}$. We may then calculate the retarded correlation function $\chi^R_B$ 
\beq
     \chi^R_{B}(q,\omega)\equiv \frac{-i\theta(t)}{L}\int dt e^{-i\omega t}\sum_{ \sigma,a}\langle [\rho_{ qa\sigma}(t),\rho_{ -qa\sigma}(0)]\rangle_B,\nonumber\\
\eeq
from Eq. \eqref{eq:Hamiltonian}, the current Hamiltonian. Here $\langle A\rangle_B \equiv \bra{0}A\ket{0}_B$. Since $\rho_{ qa \sigma}=\rho^{\xi \xi}_{ qa\sigma}+\rho^{\xi \eta}_{ qa\sigma}+\rho^{\eta \xi}_{ qa\sigma}$. The charge susceptibility now has two contributions, $\chi^{\text{Direct}}_B = \braket{[\rho^{\xi \xi},\rho^{\xi \xi}]}$ and $\chi^{\text{Mix}}_B = \braket{[\rho^{\xi \eta},\rho^{\eta \xi}]}+\braket{[\rho^{\eta \xi},\rho^{\xi \eta}]}$. Following Appendix~C, we find that
 \begin{align}
    \chi^R_{B}(q,\omega)=\frac{\rho U}{\omega^2- U^2} +\frac{v_Fq^2}{\pi \omega^2}+O\left(q^2\right). \label{eq:chi_final}
\end{align}

Here we compare the density-density correlation functions computed from the currents, Eq.\eqref{eq:chi_final}, with that obtained from the HK Hamiltonian, Eq.\eqref{eq:chi_HK}. Eq. \eqref{eq:chi_final}  captures the behavior of charge susceptibility as $q \to 0$ of the band HK model computed in \cite{MaPRB2025}. The difference between the band HK Eq. \eqref{eq:starting_point} and the current algebra Eq. \eqref{eq:Hamiltonian} is the $|q|$ contribution. In the Dirac limit where we take $k_F, m \to \infty$ while keeping $v_F = \frac{k_F}{m} = \text{Const}$, when $U$ is large, $|\chi_B - \chi|/\chi \simeq \frac{q}{k_F},\frac{\omega^2}{U v_F} \to 0$.
Thus, despite being local, our current algebra Hamiltonian captures the IR behavior $(\omega ,q \to 0)$ of the HK model at the Dirac limit. 

At last, we  address the issue of how to couple the current algebra Hamiltonian to the $U(1)$ gauge field $A_{\mu}$. This is easiest to do in the compact boson representation of $\mathfrak{su}_1(2)$ algebra, where $J_z \sim i \partial \phi$, $J^{\pm} \sim e^{\pm i\sqrt{ 2} \phi}$ \cite{cftpdf}. The minimal coupling procedure with the gauge field is $\epsilon^{\mu \nu}\partial_{\mu} \phi A_{\mu}$, with $j^{\mu}=\frac{ \delta S}{\delta A_{\mu}}=\epsilon^{\mu \nu}\partial_{\nu} \phi$, which satisfies $\partial ^{\mu}j_{\mu} = 0$. The total charge of the temporal component $Q = \int dxj^0(x) = \int dx \partial _x \phi(x) = \int dx J^z(x)$ is proportional to
\beq
    \int dx J^z(x)=\varrho^z_{q=0}=\sum_{k,a=L,R,\sigma}n_{ka\sigma}(1-n_{ka\bar{\sigma}})=N_L.
\eeq
In the hole-doping case, since there is no double occupancy in equilibrium, $N_D = 0$ and $N_L$ equal the total charge. Thus, the longitudinal part of our response  coincides with the electromagnetic response of the band HK model. Moreover, since in 1D, the transverse part is automatically zero, the formulation here yields the correct electromagnetic response.

In conclusion, we have shown that in the 1D band HK model, due to strong repulsive interactions, the propagating degrees of freedom are currents $\rho^{\alpha \beta}_{ qa\sigma}$. These currents form a $\mathfrak{su}_1(2)$ affine Lie algebra, and the low-energy, long-wavelength dynamics of the band HK model are captured by a local Hamiltonian Eq. \eqref{eq:Hamiltonian} in terms of real-space currents.  Moreover, the BJL prescription, which we used to derive the current algebra, could be applied to strongly correlated systems and is not restricted to one dimension. Further directions include the generalization of the current algebra to the Momentum Mixing HK model \cite{OHK, mai2024new}, higher-dimensional cases, or even the Hubbard model. Moreover, exploring the relationship between the WZW term and the multivaluedness of the Luttinger-Ward functional\cite{luttingerlanave} in HK models remains to be completed.

\textbf{Acknowledgements:}P. W. Phillips thanks his late colleague Rob Leigh for formative discussions on currents and their disconnect with the underlying partriculate excitations in strongly correlated systems.  In addition, we thank Gabriele La Nave and Michael Stone for instructive, characteristically level-headed remarks on currents and Barry Bradlyn, Sounak Sinha, Gaurav Tenkila, and Yuhao Ma for clarifying discussions.

\bibliographystyle{unsrt}
\bibliography{mottbib}

\appendix

\section{A. Derivation of BJL Prescription}\label{Appendix_A}

Here we present the proof of the BJL prescription \cite{bjorken1966applications,johnson1966current,treiman2014current}. For two arbitrary  states, $\ket{a}$,$\ket{b}$, the BJL prescription is
\beq
    T(\omega)&=& \int dt e^{i \omega t} \bra{a}TA(t)B(0)\ket{b} \nonumber \\
    \lim_{\omega \to \infty} \omega T(\omega)&=&i \bra{a}[A(0),B(0)]_{-} \ket{b}.
\eeq
for any two operators, $A$ and $B$.  
We may first absorb the $\omega$ factor via an integration by parts, 
\begin{align}
    \omega T(\omega)= \int _{-\infty}^{+\infty}dt(-i)\partial_{t}e^{i \omega t}\bra{a}TA(t)B(0)\ket{b} \nonumber \\
    =i\int_{- \infty}^{+\infty}dt e^{i \omega t}\partial_t \bra{a}TA(t)B(0)\ket{b}.
\end{align}
The time derivative of $TA(t)B(0)$ consists of two parts, 
\begin{align}
        T\dot{A}(t)B(0) , \nonumber
\end{align}
the smooth part,  due to the time derivative of $A(t)$, and the singular part
\begin{align}
    \delta(t)[A(t)B(0)-B(0)A(t)] \nonumber
\end{align}
generated from the time derivative of $\theta(\pm t)$.

In the limit $\omega \to \infty$, 
\begin{align}
    \lim_{\omega \to \infty}\int _{-\infty}^{+\infty}dte^{i \omega t}\bra{a}T \dot{A}(t)B(0)\ket{b}=0,
\end{align}
because of the Lebesgue theorem. Only the singular part survives the $\omega \to \infty$ limit, 
\begin{align}
    \lim_{\omega \to \infty} \int_{- \infty}^{+ \infty}dt \delta(t) e^{i \omega t} \bra{a}[A(t),B(0)]_{-}\ket{b}= \bra{a}[A(0),B(0)]_{-}\ket{b},
\end{align}
thereby completing the proof of the BJL prescription.

\section{B. Calculation of $T^{\alpha \beta ; \gamma \delta}_{\mu \nu,\sigma}$}

Here we present details of the calculation of $T^{\xi \xi;\xi \xi}_{\mu \nu\sigma}(q,\omega)$ and $T^{\xi \eta; \eta \xi}_{\mu \nu\sigma}(q,\omega)$. We omit the spin index $\sigma$ in this Appendix since these results are spin-independent. In all of these calculations, we adopt the following definition of the current operators $\rho^{\alpha \beta}_{\mu{ q \sigma}}=(\rho^{\alpha \beta}_{ q \sigma},(\rho^P)^{\alpha \beta}_{ q \sigma})$:
\begin{align}
    \rho^{\alpha \beta}_{ q \sigma}&=\rho^{\alpha \beta}_{ qL \sigma}+\rho^{\alpha \beta}_{ qR\sigma}=\sum_{ k}c^{\alpha \dagger}_{ k+q \sigma}c^{\beta}_{ k \sigma} \\
    (\rho^P)^{\alpha \beta}_{ q \sigma}&=\rho^{\alpha \beta}_{ q L \sigma}-\rho^{\alpha \beta}_{ qR\sigma}=k_F\sum_{ k} \text{sgn}(k)c^{\alpha \dagger}_{ k+q \sigma}c^{\beta}_{ k \sigma}.
\end{align}
In general, these operators should be regularized using the standard field theory approach through the addition of a cutoff $f_{\Lambda}(k^2-k_F^2)$ to project out states deep inside the filling sea. The cutoff $f_{\Lambda}(k^2-k_F^2)$ restricts the summation over k to the range $[\pm k_F - \Lambda , \pm k_F + \Lambda]$. However, since we are working with a UV finite model, it is unnecessary to regularize these operators to obtain the correct current algebra. For later convenience, we redefine $\rho^P$ by $\rho^P/k_F$.

We start with $T^{\xi \xi; \xi \xi}$ in Eq. \eqref{eq:T_factor}. For $\mu = \nu = 0$,
\begin{widetext}
    \beq
    T^{\xi \xi ; \xi \xi}_{00}(q,\omega)
    &=&\int dt \sum_{ k_1 k_2}\delta_{ k_2, k_1+q}e^{i \left( \omega + \frac{(k_1+q)^2}{2m}-\frac{k_1^2}{2m}\right)t}[\theta(t) \braket{c^{\xi \dagger}_{ k_1 + q \sigma}c^{\xi}_{ k_1 \sigma}c^{\xi \dagger}_{ k_2-q \sigma}c^{\xi}_{ k_2 \sigma}}+\theta(-t)\braket{c^{\xi \dagger}_{ k_2-q \sigma}c^{\xi}_{ k_2 \sigma}c^{\xi \dagger}_{ k_1 + q \sigma}c^{\xi}_{ k_1 \sigma}}] \nonumber \\
    &=&\frac{1}{2}\int dt \sum_{ k_1,k_2} e^{i \left( \omega + \frac{(k_1+q)^2}{2m}-\frac{k_1^2}{2m} \right)t}\delta_{k_2,k_1+q}[\theta(t)\braket{n^S_{ k_2 }}\braket{n^O_{ k_1 }}+ \theta(-t)\braket{n^O_{ k_2 }}\braket{n^S_{ k_1 }}] \nonumber \\
    &=&\frac{1}{2}\theta(q)\left[\sum_{k_1 = -k_F -q}^{-k_F}\frac{1}{-i\left( \omega + \frac{k_1q}{m}+\frac{q^2}{2m}\right)+0^+} + \sum_{k_1 = k_F-q}^{k_F}\frac{1}{i\left( \omega + \frac{k_1q}{m}+\frac{q^2}{2m}\right)+0^+} \right] \nonumber \\
    &+&\frac{1}{2}\theta(-q)\left[\sum_{k_1 = k_F}^{k_F-q}\frac{1}{-i\left( \omega + \frac{k_1q}{m}+\frac{q^2}{2m}\right)+0^+} + \sum_{k_1 = -k_F}^{-k_F-q}\frac{1}{i\left( \omega + \frac{k_1q}{m}+\frac{q^2}{2m}\right)+0^+} \right].
\eeq
\end{widetext}

To extract the equal-time commutator, we only need to evaluate the residue of $\omega^{-1}$. Since in the $\omega \to \infty$ limit there is no singularity in the integrand, we may expand the integrand in terms of $\omega^{-n}$ before doing the integration. When $q>0$, we find that 
\beq
    T^{\xi \xi; \xi \xi}_{00}(q,\omega) \simeq \sum_{k_1 = -k_F-q}^{-k_F}\frac{1}{-i \omega}+\sum_{k_1 = k_F-q}^{k_F}\frac{1}{i \omega}  = 0.
\eeq
This yields $T^{\xi \xi; \xi \xi}_{00}(\omega)=0$. The same result holds for the $q<0$ case. Calculation of $T^{\xi \xi; \xi \xi}_{11}$ is rather similar, which also yields 0. For $\mu = 0, \nu =1$, we have
\begin{widetext}
    \beq
    T^{\xi \xi; \xi \xi}_{01}(q,\omega) 
    &=&\int dt \sum_{k_1 k_2}\delta_{ k_2, k_1+q}e^{i \left( \omega +\frac{(k_1+q)^2}{2m}-\frac{k_1^2}{2m}\right)t}\text{sgn}(k_2) [\theta(t)\braket{c^{\xi \dagger}_{ k_1 + q \sigma}c^{\xi}_{ k_1 \sigma}c^{\xi \dagger}_{ k_2-q \sigma}c^{\xi}_{ k_2 \sigma}}+\theta(-t)\braket{c^{\xi \dagger}_{ k_2-q \sigma}c^{\xi}_{ k_2 \sigma}c^{\xi \dagger}_{ k_1 + q \sigma}c^{\xi}_{ k_1 \sigma}}] \nonumber \\
    &=&\frac{1}{2}\int dt \sum_{ k_1,k_2} e^{i \left( \omega + \frac{(k_1+q)^2}{2m}-\frac{k_1^2}{2m} \right)t}\delta_{k_2,k_1+q}\text{sgn}(k_2)[\theta(t)\braket{n^S_{ k_2 }}\braket{n^O_{ k_1 \sigma}}+ \theta(-t)\braket{n^O_{ k_2 }}\braket{n^S_{ k_1 \sigma}}] \nonumber \\
    &=&\frac{\theta(q)}{2}\left[\sum_{k_1 = -k_F -q}^{-k_F}\frac{\text{sgn}(k_1+q)}{-i\left( \omega + \frac{k_1q}{m}+\frac{q^2}{2m}\right)+0^+} + \sum_{k_1 = k_F-q}^{k_F}\frac{\text{sgn}(k_1+q)}{i\left( \omega + \frac{k_1q}{m}+\frac{q^2}{2m}\right)+0^+} \right] \nonumber \\
    &+&\frac{\theta(-q)}{2}\left[\sum_{k_1 = k_F}^{k_F-q}\frac{\text{sgn}(k_1+q)}{-i\left( \omega + \frac{k_1q}{m}+\frac{q^2}{2m}\right)+0^+} + \sum_{k_1 = -k_F}^{-k_F-q}\frac{\text{sgn}(k_1+q)}{i\left( \omega + \frac{k_1q}{m}+\frac{q^2}{2m}\right)+0^+} \right].
\eeq 
\end{widetext}

In the limit $\omega \to \infty, q \to 0$ , for $q>0$ case, the contribution from the $\omega^{-1}$ pole is
\begin{align}
    T^{\text{LL;LL}}_{01}(q,\omega)\simeq\frac{1}{2}\left( \sum_{k_1= -k_F-q}^{-k_F} \frac{-1}{-i \omega}+\sum_{k_1 = k_F-q}^{k_F}\frac{1}{i \omega} \right)=\frac{L q}{2 \pi i \omega}.
\end{align} For $q<0$ case, it yields the same result.

For the $T^{\xi \eta; \eta \xi}$ case, the $\theta(-t)$ part does not contribute. Since there is no double occupancy in the ground state, $\braket{\rho^{\eta \xi}\rho^{\xi \eta}}=0$. We also set $\mu = \nu = 0$ to obtain
\begin{widetext}
    \beq
    T^{\xi \eta; \eta \xi}_{00}(q,\omega) \nonumber 
    &=&\int dt \sum_{ k_1 k_2}\delta_{ k_2, k_1+q}e^{i \left( \omega-U + \frac{k_1q}{m}+\frac{q^2}{2m} \right)t}\theta(t) \braket{c^{\xi \dagger}_{ k_1 + q \sigma}c^{\eta}_{ k_1 \sigma}c^{\eta \dagger}_{ k_2-q \sigma}c^{\xi}_{ k_2 \sigma}} \nonumber \\
    &=&\frac{1}{4}\sum_{ k_1, k_2}\delta_{ k_2, k_1 + q}\frac{ \braket{n^S_{ k_1}}\braket{n^S_{ k_2}}}{-i \left(\omega - U + \frac{k_1q}{m}+\frac{q^2}{2m} \right) + 0^+} \nonumber \\
    &=&\frac{L \theta(q)}{8 \pi} \int_{-k_F}^{k_F-q}dk \frac{1}{-i \left(\omega - U + \frac{k_1 q}{m}+\frac{q^2}{2m} \right) + 0^+} + \frac{L \theta(-q)}{8 \pi} \int_{-k_F-q}^{k_F}dk \frac{1}{-i \left( \omega - U + \frac{k_1 q}{m}+\frac{q^2}{2m} \right) + 0^+} .
\eeq
\end{widetext}
In the limit $\omega\rightarrow \infty$, we find the leading term , 
\beq
    T^{\xi \eta; \eta \xi}_{00}(q,\omega) &=& \frac{L \theta(q)}{8 \pi}\frac{2k_F-q}{-i \omega} + \frac{L \theta(-q)}{8 \pi}\frac{2k_F + q}{-i \omega} \nonumber \\
    &=& \frac{L}{8 \pi}\frac{2k_F - |q|}{-i \omega},
\eeq
to be linear in $|q|$. The linear dependence arises from the difference in momenta for the occupied $n_{k\sigma}$ and unoccupied $1-n_{k\sigma}$ states.

Since $L \rho = N$, we have $-i\omega T^{\xi \eta; \eta \xi}_{00}(q,\omega)=\frac{N}{4}-\frac{|q|L}{8 \pi}$. Computation of $T^{\xi \eta; \eta \xi}_{11}(q,\omega)$ yields identical results.
For $T^{\xi \eta ; \eta \xi}_{01}(q, \omega)$, we find that
\begin{widetext}
\beq
    T^{\xi \eta ; \eta \xi}_{01}(q, \omega) 
    &=&\int dt \sum_{ k_1 k_2}\delta_{ k_2, k_1+q}e^{i \left( \omega - U + \frac{k_1q}{m}+\frac{q^2}{2m} \right)t}\theta(t)\text{sgn}(k_2) \braket{c^{\xi \dagger}_{ k_1 + q \sigma}c^{\eta}_{ k_1 \sigma}c^{\eta \dagger}_{ k_2-q \sigma}c^{\xi}_{ k_2 \sigma}} \nonumber \\
    &=&\frac{1}{4}\sum_{ k_1, k_2}\delta_{ k_2, k_1 + q}\text{sgn}(k_2)\frac{ \braket{n^S_{ k_1}}\braket{n^S_{ k_2}}}{-i \left(\omega - U + \frac{k_1q}{m}+\frac{q^2}{2m} \right) + 0^+} \nonumber \\
    &=&\frac{L \theta(q)}{8 \pi} \int_{-k_F}^{k_F-q}dk_1  \frac{\text{sgn}(k_1+q)}{-i \left(\omega - U + \frac{k_1 q}{m}+\frac{q^2}{2m} \right) + 0^+} + \frac{L \theta(-q)}{8 \pi} \int_{-k_F-q}^{k_F}dk_1 \frac{\text{sgn}(k_1+q)}{-i \left( \omega - U + \frac{k_1 q}{m}+\frac{q^2}{2m} \right) + 0^+} .
\eeq
\end{widetext}
Thus, the infinite frequency limit is
\beq
    T^{\xi \eta ; \eta \xi}_{01}(q,\omega)= \frac{L  q}{-i8 \pi \omega}+O(q^2).
\eeq

\section{C. Calculation of Charge Susceptibility from the Current Algebra}

In this section, we calculate $\braket{\rho \rho}^{\text{Direct}}_B$ and $\braket{\rho \rho}^{\text{Mix}}_B$. We start with $\braket{\rho \rho}^{\text{Direct}}_B(q,\omega)$. 
\begin{widetext}
    \beq
    \braket{\rho \rho}_{B}^{\text{Direct}}(q,\omega)&=&-\frac{i}{L}\int_0^{\infty} dt e^{-i\omega t}\sum_{ \sigma}[e^{iv_Fqt}\langle [\rho^{\xi \xi}_{ qR\sigma},\rho^{\xi \xi}_{ -qR\sigma}] \rangle_B+e^{-iv_Fqt}\langle  [\rho^{\xi \xi}_{ qL\sigma},\rho^{\xi \xi}_{ -qL\sigma}]\rangle_B] \nonumber \\
    &=&-\frac{1}{L}\sum_{\sigma}\left(\frac{\langle [\rho^{\xi \xi}_{ qR\sigma},\rho^{\xi \xi}_{ -qR\sigma}] \rangle_B}{\omega - v_Fq-i0^+}+\frac{\langle  [\rho^{\xi \xi}_{ qL\sigma},\rho^{\xi \xi}_{ -qL\sigma}]\rangle_B}{\omega + v_Fq-i0^+} \right).
\eeq
\end{widetext}
The spectral weights are $\langle [\rho^{\xi \xi}_{ qR\sigma},\rho^{\xi \xi}_{ -qR\sigma}] \rangle_B=-\frac{qL}{4\pi}$ and $\langle  [\rho^{\xi \xi}_{ qL\sigma},\rho^{\xi \xi}_{ -qL\sigma}]\rangle_B=\frac{qL}{4\pi}$. In the limit $q \to 0$, the leading order terms are 
\beq
    \braket{\rho \rho}_{B}^{\text{Direct}}(q,\omega)&=&\sum_{\sigma}\frac{q}{4\pi}\left( \frac{1}{\omega - v_Fq}-\frac{1}{\omega + v_F q} \right) \nonumber \\
    &=&\frac{v_Fq^2}{\pi \omega^2}+O(q^2).
\eeq

We next calculate $\braket{\rho \rho}_B^{\text{Mix}}(q,\omega)$ using the same method. In terms of the commutator, we find that
\begin{widetext}
    \beq
    \braket{\rho \rho}^{\text{Mix}}_{B}(q,\omega)
    &=&-\frac{1}{L}\sum_{ \sigma }\left[ \frac{\bra{0} [\rho^{\xi \eta}_{ qR\sigma},\rho^{\eta \xi}_{ -qR\sigma}]\ket{0}}{(\omega-v_Fq+U)-i0^{+}} + \frac{\bra{0} [\rho^{\xi \eta}_{ qL\sigma},\rho^{\eta \xi}_{ -qL\sigma}]\ket{0}}{(\omega+v_Fq+U)-i0^+}+ \frac{\bra{0} [\rho^{\eta \xi}_{ qR\sigma},\rho^{\xi \eta}_{ -qR\sigma}]\ket{0}}{(\omega-v_Fq-U)-i0^{+}} + \frac{\bra{0} [\rho^{\eta \xi}_{ qL\sigma},\rho^{\xi \eta}_{ -qL\sigma}]\ket{0}}{(\omega+v_Fq-U)-i0^+}\right] \nonumber \\
    &=&-\sum_{\sigma} \left[\left( \frac{\rho}{8}-\frac{q}{8\pi} \right)\left( \frac{1}{\omega-v_Fq + U}- \frac{1}{\omega+v_Fq-U}\right) + \left( \frac{\rho}{8}+\frac{q}{8\pi} \right)\left( \frac{1}{\omega+v_Fq+U}- \frac{1}{\omega-v_Fq-U}\right) \right] \nonumber \\
    &=&\frac{ \rho U}{\omega^2 - U^2}+O(q^2) .\label{eq:ChiMix}
\eeq
\end{widetext}
Consequently, we obtain the expression Eq.\eqref{eq:chi_final} through adding $\braket{\rho \rho}^{\text{Direct}}_B$ and $\braket{\rho \rho}^{\text{Mix}}_B$.  This corroborates that the formulation in terms of local currents and the HK Hamiltonian in terms of fermionic operators are in agreement in the Dirac limit.

\end{document}